\documentclass{osa-article}

\journal{osajournal}

\articletype{Research Article}

		\usepackage{upgreek}

		\catcode`@=11
		\def\nvphantom{\v@true\h@false\nph@nt}
		\def\nhphantom{\v@false\h@true\nph@nt}
		\MakeRobust{\vphantom}
		\def\nphantom{\v@true\h@true\nph@nt}
		\def\nph@nt{\ifmmode\def\next{\mathpalette\nmathph@nt}\else\let\next\nmakeph@nt\fi\next}
		\def\nmakeph@nt#1{\setbox\z@\hbox{#1}\nfinph@nt}
		\def\nmathph@nt#1#2{\setbox\z@\hbox{$\m@th#1{#2}$}\nfinph@nt}
		\def\nfinph@nt{\setbox\tw@\null \ifv@ \ht\tw@\ht\z@ \dp\tw@\dp\z@\fi\ifh@ \wd\tw@-\wd\z@\fi \box\tw@}
		\usepackage{stackengine}[2013-10-15]
		\usepackage{scalerel}
		
		\renewcommand{\log}[1]{\text{ln}\!\left({#1}\right)}

		\newcommand{\Real}[1]{\text{Re}\!\left({#1}\right)}

		\newcommand{\Imag}[1]{\text{Im}\!\left({#1}\right)}

		\newcommand\limit[1]{\mathop{\vcenter{\hbox{$\lim\limits_{#1}$}}}\,}
		\newcommand{\diff}[1]{\text{d}{#1}}

		\newcommand{\integral}[4]{\int_{#1}^{#2}{#3}\ifthenelse{\isempty{#4}}{}{\,\text{d}{#4}}}
		\newcommand{\ointegral}[3]{\oint_{#1}{#2}\ifthenelse{\isempty{#3}}{}{\,\text{d}{#3}}}
		\usepackage{scalerel}[2016-12-29]

		\usepackage[makeroom]{cancel}
		
		\renewcommand{\leq}{\leqslant}
		\renewcommand{\geq}{\geqslant}
		
		\renewcommand{\.}{\,\!}
		\newcommand*{\eq}[1]{\begin{eqnarray}#1\end{eqnarray}}
		
		\usepackage{etoolbox}
		\AtBeginEnvironment{eqnarray}{\setlength{\arraycolsep}{2pt}}

		\newcommand{\figwidth}{1\linewidth}
		
		\newcommand{\graphwidth}{1.2*\figwidth/sqrt(2)}
		
		\usepackage{pgfplots}
		\pgfplotsset{compat=1.9}
		\usepgfplotslibrary{fillbetween}
		\definecolor{blood}{rgb}{0.8,0,0}
		\definecolor{sodium}{rgb}{1,0.75,0}
		\newcommand*{\fig}[4]{\begin{figure}[b!]\begin{center}{\includegraphics[width=#2]{#1}}\captionsetup{singlelinecheck=off}\end{center}\caption[.]{#3}\label{#4}\end{figure}}
		\newcommand*{\figt}[4]{\begin{figure}[t!]\begin{center}{\includegraphics[width=#2]{#1}}\captionsetup{singlelinecheck=off}\end{center}\caption[.]{#3}\label{#4}\end{figure}}

		\usepackage{xifthen}
		\newcommand{\article}[6]{#1 (#2): \textit{#3}, #4 #5, \mbox{#6}}

		\newcommand{\book}[6]{#1 (#2): \textit{#3}, \ifthenelse{\isempty{#4}}{}{\ifthenelse{\equal{\detokenize{#4}}{\detokenize{2}}}{second}{\ifthenelse{\equal{\detokenize{#4}}{\detokenize{3}}}{third}{\ifthenelse{\equal{\detokenize{#4}}{\detokenize{4}}}{fourth}{\ifthenelse{\equal{\detokenize{#4}}{\detokenize{5}}}{fifth}{\ifthenelse{\equal{\detokenize{#4}}{\detokenize{6}}}{sixth}{\ifthenelse{\equal{\detokenize{#4}}{\detokenize{7}}}{seventh}{\ifthenelse{\equal{\detokenize{#4}}{\detokenize{8}}}{eighth}{\ifthenelse{\equal{\detokenize{#4}}{\detokenize{9}}}{ninth}{#4}}}}}}}} edition, }#5\ifthenelse{\isempty{#6}}{}{, \mbox{#6}}}
		\newcommand{\bookarticle}[7]{#1 (#2): \textit{#3}, in \textit{#4}\ifthenelse{\isempty{#5}}{}{ #5}, #6, \mbox{#7}}

		\newcommand{\Ec}{\vec{E}^\text{\,c}}
		\newcommand{\Eco}{E^\text{\,c}_0}
		\newcommand{\Ee}{\vec{E}^\text{\,e}}
		\newcommand{\Eeo}{E^\text{\,e}_0}
		\newcommand{\Ei}{\vec{E}^\text{\,i}}
		\newcommand{\Eio}{E^\text{\,i}_0}
		\newcommand{\Es}{\vec{E}^\text{\,s}}
		\newcommand{\Eso}{E^\text{\,s}_0}

		\newcommand{\FEs}{\vec{\mathcal{E}}^\text{\,s}}

		\newcommand{\FEo}{\vec{\mathcal{E}}^{\,0}}
		
		\newcommand{\FEox}{\mathcal{E}^{\,0}_x}
		\newcommand{\FEoy}{\mathcal{E}^{\,0}_y}
		\newcommand{\FEoz}{\mathcal{E}^{\,0}_z}
		\DeclareRobustCommand{\hbar}{\mathrlap{\mspace{-1mu}\bar{\phantom{x}}}h}
		\newcommand{\keff}{k_\text{e}}
		\newcommand{\km}{k_\text{m}}
		\newcommand{\kx}{k_x}
		\newcommand{\ky}{k_y}
		\newcommand{\kp}{k_\text{p}}
		\newcommand{\neff}{n_\text{e}}
		\newcommand{\nm}{n_\text{m}}
		\newcommand{\np}{n_\text{p}}

		\newcommand{\evalr}{(\vec{r})}

		\newcommand{\Sfwd}{S(0)}
		
		\newcommand{\Xm}{X_\text{m}}
		
		\newcommand{\um}{$\upmu$m}
		\newcommand{\ummm}{\um$^3$}
		\newcommand{\mmmu}{\um$^{-3}$}

\begin{document}\sloppy

\twocolumn[{%
\begin{@twocolumnfalse}

\title{An unambiguous derivation of the effective refractive index of biological suspensions and an extension to dense tissue such as blood}

\author{Alexander Nahmad-Rohen\authormark{1,*} and Augusto Garc\'ia-Valenzuela\authormark{1}}

\address{\authormark{1}Instituto de Ciencias Aplicadas y Tecnolog\'ia, Universidad Nacional Aut\'onoma de M\'exico, Circuito Exterior S/N, Avenida Universidad 3000, Coyoac\'an, Mexico City, Mexico}

\email{\authormark{*}alexander.nahmad@icat.unam.mx} 



\begin{abstract}
The van de Hulst formula provides an expression for the effective refractive index or effective propagation constant of a suspension of particles of arbitrary shape, size and refractive index in an optically homogeneous medium. However, its validity for biological matter, which often consists of very dense suspensions of cells, is unclear because existing derivations of the formula or similar results rely on far-field scattering and/or on the suspension in question being dilute. We present a derivation of the van de Hulst formula valid for suspensions of large, tenuous scatterers ---the type biological suspensions are typically made of--- which does not rely on these conditions, showing that they are not strictly necessary for the formula to be valid. We apply these results specifically to blood and epithelial tissue. Furthermore, we determine the true condition for the formula to be valid for these types of tissues. We finally provide a simple way to estimate ---and, more importantly, correct--- the error incurred by the van de Hulst formula when this condition is not met.
\end{abstract}

\end{@twocolumnfalse}}]


\section{Introduction}\label{sec-intro}

The physical study of biological matter is inherently complex. Even a relatively simple medium such as blood, which consists of a suspension of various different cell types (the most numerous of which has a very simple structure compared to most cells) in an optically isotropic medium, presents challenges which are not always easy to address.

Optics carries the unique benefit of sometimes (depending on the particular technique employed) being able to probe translucent systems without altering them. The introduction of mechanical or chemical probes used in standard blood tests is invasive and potentially destructive. Conversely, visible and infrared light are harmless to a blood sample and compatible with noninvasive probing.

The principal difficulty in the optical study of biological materials stems from the high extinction coefficient they typically have. This extinction coefficient is equal to the sum of the scattering and absorption coefficients. Of the two, scattering is stronger in this type of material due to a highly inhomogeneous composition. Therefore, a complete optical study of this type of material must begin by properly understanding how the material scatters light.

It is sometimes appropriate to look at a heterogeneous material as though it were a homogeneous one. When this is done, the underlying theory is called an effective-medium theory. Average properties, such as an effective refractive index, are calculated as functions of the properties of the material's constituents; they therefore can, if one is careful, be useful quantities to study the material in question. For example, the effective refractive index of a heterogeneous material correctly determines the direction of refraction of the coherent component of light after passing through the material, as well as the Fresnel transmission coefficient for coherent light\cite{ref-Foldy-PR67}, but it fails to correctly describe reflection at the material's edge or describe the behaviour of the diffuse component of light\cite{ref-Mohammadi-ACIS62}. This is because effective-medium theories deal only with the average effect of the material on light. To take reflection as an example, light that is reflected at the material's edge only interacts with a small portion of the sample, so the effect of that portion of the sample on that light is not the same as the average effect over a distance of many wavelengths on light which travels through the sample\cite{ref-GarciaValenzuela-OE16}. Despite the caution one must exercise when working with effective quantities, they have important applications, such as medical diagnosis\cite{ref-NahmadRohen-PS91}.

One of the better-known models compatible with materials of the biologial type ---formed by large (compared to the wavelength), arbitrarily shaped, arbitrarily oriented particles--- is that of H~C~van~de~Hulst, developed in the mid-1900s\cite{ref-vdH-LSSP}. This has been shown to yield accurate results for large tenuous particles (see below), even sometimes for high concentrations of particles\cite{ref-Alexander-JCSFT277,ref-Meeten-OC134,ref-Meeten-MST8}. Unfortunately, the derivations that are presented in van de Hulst's book and in other important texts on optical scattering seem to imply that the model requires a very dilute suspension\cite{ref-vdH-LSSP,ref-Barrera-JOSAA20,ref-Barrera-PRB75,ref-GutierrezReyes-PSSb249} (which is very often not the case with biological materials) and/or an observation point far from the material being studied\cite{ref-Foldy-PR67,ref-vdH-LSSP,ref-Bohren-ASLSP,ref-Lax-RMP23}. Other important derivations of the van de Hulst formula or similar results\cite{ref-Twersky-JOSA52,ref-Twersky-JMP3,ref-Twersky-JOSA60a} employ the Wentzel-Kramers-Brillouin (WKB) approximation, which is a variation of the Rayleigh-Gans (or sometimes Rayleigh-Gans-Debye) approximation to take into account phase retardation by the particles in the suspension. The WKB approximation is similar to van de Hulst's anomalous-diffraction approximation\cite{ref-vdH-LSSP} (which we use in the present work) and is, in fact, more or less equivalent to it in the sense that the two approximations yield the same effective propagation constant except for a multiplicative factor $(\np/\nm+1)/2$, where $\np$ is the refractive index of the particles and $\nm$ is that of the surrounding medium\cite{ref-Klett-AO31}, a factor which is approximately 1 for tenuous particles (i.e.~particles with $|\np/\nm-1|\ll1$). However, these derivations require either extremely tenuous particles or ones that are smaller than the wavelength, as imposed by the condition that $k(\np-\nm)b\ll1$, where $k$ is the wave number and $b$ is the distance travelled by the light within a particle\cite{ref-Twersky-JMP3}; for visible light and cells, this quantity is most certainly not much smaller than 1 (it equals between 1.1 and 17.6 in the case of an erythrocyte suspended in blood plasma, for example, depending on the erythrocyte's orientation and the wavelength).

In this work, we will show that the assumptions of a dilute suspension and far-field observation are not generally necessary for the van de Hulst model to be valid in the case of biological suspensions. We will also calculate a correction term for it. We shall henceforth assume the light interacting with the suspension is visible (i.e.~has a wavelength of 400--800~nm).

\section{General scattering theory and the effective-field approximation}\label{sec-EFA}

Consider a plane wave with electric field $\Ei=\Eio e^{i\km z}\hat{e}$ travelling in the $z$ direction through a medium formed by $N$ particles suspended in a homogeneous medium with refractive index $\nm$ and propagation constant $\km$; the unit vector $\hat{e}$ indicates the polarisation of the incident wave and is perpendicular to the direction in which the wave propagates. We will consider the case in which the number of particles is very large and there are no external factors influencing the particles' configurations, whereby we may assume two things: all the particles are randomly oriented (with all possible orientations occurring with the same probability) and randomly positioned, and the contribution of a single particle's scattered field is negligible compared to the total field (the latter assumption is notably false for particles with extremely elongated shapes, but such particles are beyond the scope of the present work).

The total field at any point $\vec{r}$ is given by
\eq{\vec{E}\evalr & = & \Ei\evalr+\sum_{j=1}^N\Es_j\evalr,}
where \smash{$\Es_j$} is the field scattered by the $j$-th particle. This scattered field is a function of the field that excites the $j$-th particle, which we denote by $\Ee_j$, and of the particle's properties. The exciting field, in turn, equals the incident field plus the fields scattered by all the other particles (figure~\ref{fig-Ee}):
\eq{\Ee_j\evalr & = & \Ei\evalr+\sum_{\ell\neq j}\Es_\ell\evalr.}

\figt{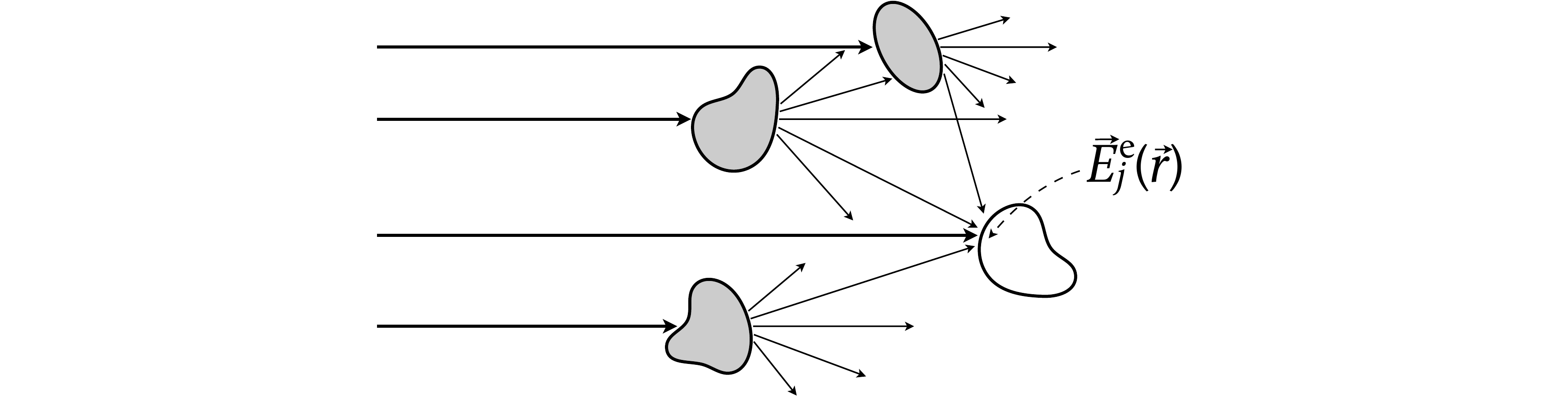}{\figwidth}{The exciting field at any point equals the sum of the incident field and the fields scattered by all the particles.}{fig-Ee}

The average of the total field over all possible configurations (positions and orientations) of the collection of particles, which we denote by $\Ec\equiv\langle\vec{E}^{\,\!}\rangle$ and call the coherent field, is
\eq{\Ec\evalr & = & \Ei\evalr+\sum_{j=1}^N\left\langle\!\Es_j\evalr\!\right\rangle.\label{eq-Ecsum}}
So far, our derivation mirrors Twersky's\cite{ref-Twersky-JOSA52}.

Suppose now that all particles are identical. If this is true, then we may replace the sum with a multiplicative factor $N$ in equation~\ref{eq-Ecsum}, obtaining
\eq{\Ec\evalr & = & \Ei\evalr+N\left\langle\!\Es_j\evalr\!\right\rangle.}
The average can be written as an integral over all of space weighted by the probability density function of finding the $j$-th particle in a given configuration. Since we are already taking the average over the particle's possible orientations (with the assumption that all orientations occur with equal probability), we need worry only about its position. Assuming the probability density governing the positions of the particles' centres is uniform within a region of volume $V$ and zero outside of this region, this probability density in the region is given by $1/V$, whereby
\eq{\Ec\evalr & = & \Ei\evalr+\rho\integral{V}{}{\Es_j\evalr}{\vec{r}_j},}
where $\rho=N/V$ is the density of particles in the suspension and $\vec{r}_j$ is the position of the $j$-th particle's centre.

All that remains, then, is to calculate \smash{$\Es_j$}, which requires knowledge of \smash{$\Ee_j$} and of how the particle affects \smash{$\Ee_j$}. To circumvent the former problem, it is customary to make the approximation \smash{$\langle\Ee_j\rangle\approx\Ec$}. This is called the effective-field approximation.

\figt{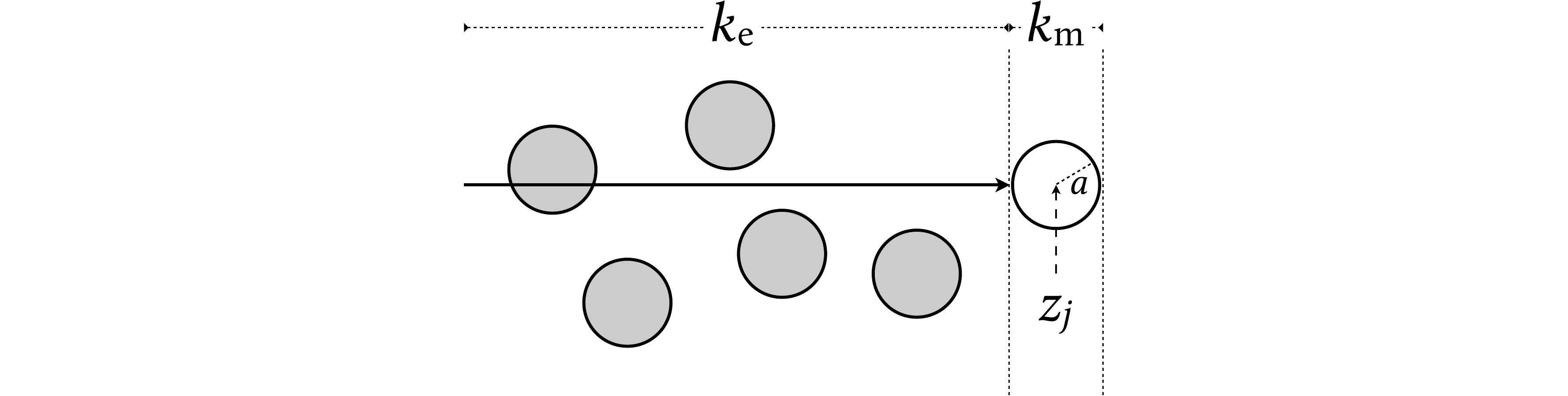}{\figwidth}{The phase of the coherent field in the vicinity of a particle needs to be separated into a component dependent on the effective propagation constant $\keff$ and a component dependent on the surrounding medium's propagation constant $\km$.}{fig-Ec}

By rotational and translational symmetry, \smash{$\Ec$} must be a plane wave travelling in the $z$ direction, just like \smash{$\Ei$}, a result already recognised in 1962 by Twersky\cite{ref-Twersky-JOSA52} (this ceases to be true if some particle orientations or positions occur with higher probabilities than others and thus one or both of these symmetries is/are broken, a case we will not consider here). In general, we can write the coherent field as $\Ec(z)=\Eco e^{i\keff z}\hat{e}$, where $\keff$ is the effective propagation constant of the suspension, from which we can define an effective refractive index $\neff=\nm\keff/\km$. In the vicinity of the $j$-th particle, however, we have
\eq{\Ec(z) & = & \Eco e^{i\left(\keff(z_j-a)+\km(z-z_j+a)\vphantom{b^2}\right)}\hat{e},}
where $a$ is the characteristic size of the particle (for example, if the particle is spherical, $a$ is its radius). This is because \smash{$\keff$} refers to the effective propagation of light through the suspension over large (compared to the wavelength) distances and has no validity locally\cite{ref-GarciaValenzuela-OE16,ref-Barrera-PRB75,ref-GutierrezReyes-PSSb249,ref-GutierrezReyes-JPCB118}, whereby for the local field in the particle's vicinity we must take the propagation constant of the medium surrounding the particles (figure~\ref{fig-Ec}).

\section{Anomalous diffraction}\label{sec-anomdiff}

Certain types of suspension which are of practical interest (for example, biological tissue) are amenable to approximations which allow a fairly simple computation of $\Es_j$. One such approximation, called anomalous diffraction\cite{ref-vdH-LSSP}, assumes the only effect of a particle on the exciting field is to introduce a phase shift $\beta=(\kp-\km)b$, where $\kp$ is the propagation constant inside the particle and $b=b(x,y)$ is the distance travelled by the exciting field through the particle (figure~\ref{fig-Es}). This is valid for particles which are much larger than the wavelength of the exciting field (so they may be considered media in their own right with a well-defined propagation constant $\kp$) and tenuous (as mentioned in section~\ref{sec-intro}, we write this condition as $|\np/\nm-1|\ll1$; this implies that refraction at the particle edges is negligible). Blood, for example, satisfies these conditions for visible light\cite{ref-NahmadRohen-PS91}, as do biological suspensions in general\cite{ref-Xu-OL30}. This is fortunate because the general case requires unwieldy mathematics even for simple geometries\cite{ref-GutierrezReyes-MPE2019}.

\figt{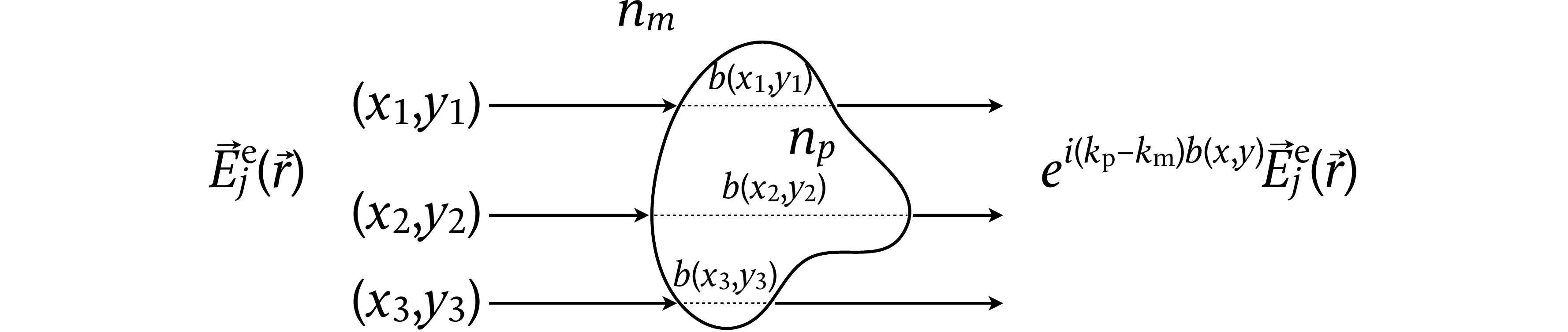}{\figwidth}{In the anomalous-diffraction approximation, the only effect of a particle on the light that passes through it is to introduce a phase factor $\beta=(\kp-\km)b(x,y)$.}{fig-Es}

\section{Derivation of the van de Hulst formula}\label{sec-vdH}

\subsection{Random suspensions}\label{subsec-vdHrandom}

Blood is a tissue formed by erythrocytes, leukocytes and thrombocytes suspended in plasma. It is a random medium, meaning the particles are randomly positioned and oriented. By far the most abundant cells present in blood are erythrocytes; all other cells combined amount to a volume fraction of about 1\%\cite{ref-Guyton-TMP,ref-Schaller-HBPP}. For this reason and in order to keep our mathematical expressions simple, we shall derive van de Hulst's formula for a medium consisting of a single particle type in this section. However, generalisation of the formula to suspensions with multiple particle types is straightforward\cite{ref-NahmadRohen-PS91}.

Using the anomalous-diffraction approximation, for $\vec{r}=(x,y,z\gtrsim z_j)$ (i.e.~immediately after the $j$-th particle) we have
\eq{\Es_j(\vec{r}) & = & \left(e^{i(\kp-\km)b}-1\right)\!\Ee_j\evalr.\label{eq-EsAD}}

To obtain the scattered field farther from the particle (say, beyond the expansion plane given by $z=z_j+a)$, we take the two-dimensional Fourier transform of \smash{$\Es_j$} at the expansion plane, given by
\eq{\FEs_j(\kx,\ky,z_j+a) & = & \frac{1}{2\pi}\integral{-\infty}{\infty}{\integral{-\infty}{\infty}{\Es_j(x,y,z_j+a)\ \times}{}}{}\nonumber\\
& & \phantom{\frac{1}{2\pi}\integral{-\infty}{\infty}{\integral{-\infty}{\infty}{}{}}{}}\times e^{-i(\kx x+\ky y)}\,\diff{x}\,\diff{y},\nonumber\\}
and then the inverse Fourier transform of this,
\eq{\Es_j(x,y,z_j+a) & = & \frac{1}{2\pi}\integral{-\infty}{\infty}{\integral{-\infty}{\infty}{\FEs_j(\kx,\ky,z_j+a)\ \times}{}}{}\nonumber\\
& & \phantom{\frac{1}{2\pi}\integral{-\infty}{\infty}{\integral{-\infty}{\infty}{}{}}{}}\times e^{i(\kx x+\ky y)}\,\diff{\kx}\,\diff{\ky}.\nonumber\\}

So far we have done nothing, but we note that, since each cartesian component of \smash{$\FEs_j$} must be a solution to the Helmholtz equation\cite{ref-Collin-ARP}, the $z$ dependence of \smash{$\FEs_j$} beyond the expansion plane must be that of a plane wave travelling in the $z$ direction:
\eq{\FEs_j(\kx,\ky,z) & = & \FEo_j(\kx,\ky)e^{i\km z}}
for $z>z_j+a$ (note that \smash{$\FEo_j$} is only a function of $\kx$ and $\ky$). Therefore, for these values of $z$, we have
\eq{\Es_j\evalr & = & \frac{1}{2\pi}\integral{-\infty}{\infty}{\integral{-\infty}{\infty}{\FEo_j(\kx,\ky)\ \times}{}}{}\nonumber\\
& & \phantom{\frac{1}{2\pi}\integral{-\infty}{\infty}{\integral{-\infty}{\infty}{}{}}{}}\times e^{i(\kx x+\ky y+\km z)}\,\diff{\kx}\,\diff{\ky}.\nonumber\\\label{eq-Helmholtz}}

For the sake of completeness, we should mention at this point that $\Es_j$ satisfies Gauss's law, \smash{$\nabla\cdot\Es_j=0$}, on both sides of which one can take the Fourier transform to obtain an expression for $\FEoz$ in terms of $\FEox$ and $\FEoy$, which, together with equation~\ref{eq-Helmholtz}, fully determines the scattered field away from the particle\cite{ref-Collin-ARP}. (Ultimately, the average of \smash{$\FEoz$} will be zero because $\Ec$ is a plane wave travelling in the $z$ direction. Note that we have written $\km$ instead of $k_z$ in equation~\ref{eq-Helmholtz} because from the beginning of section~\ref{sec-EFA} we defined $\km$ to be the wave vector of the incident light in the medium surrounding the particles and this wave vector only has a $z$ component.)

\fig{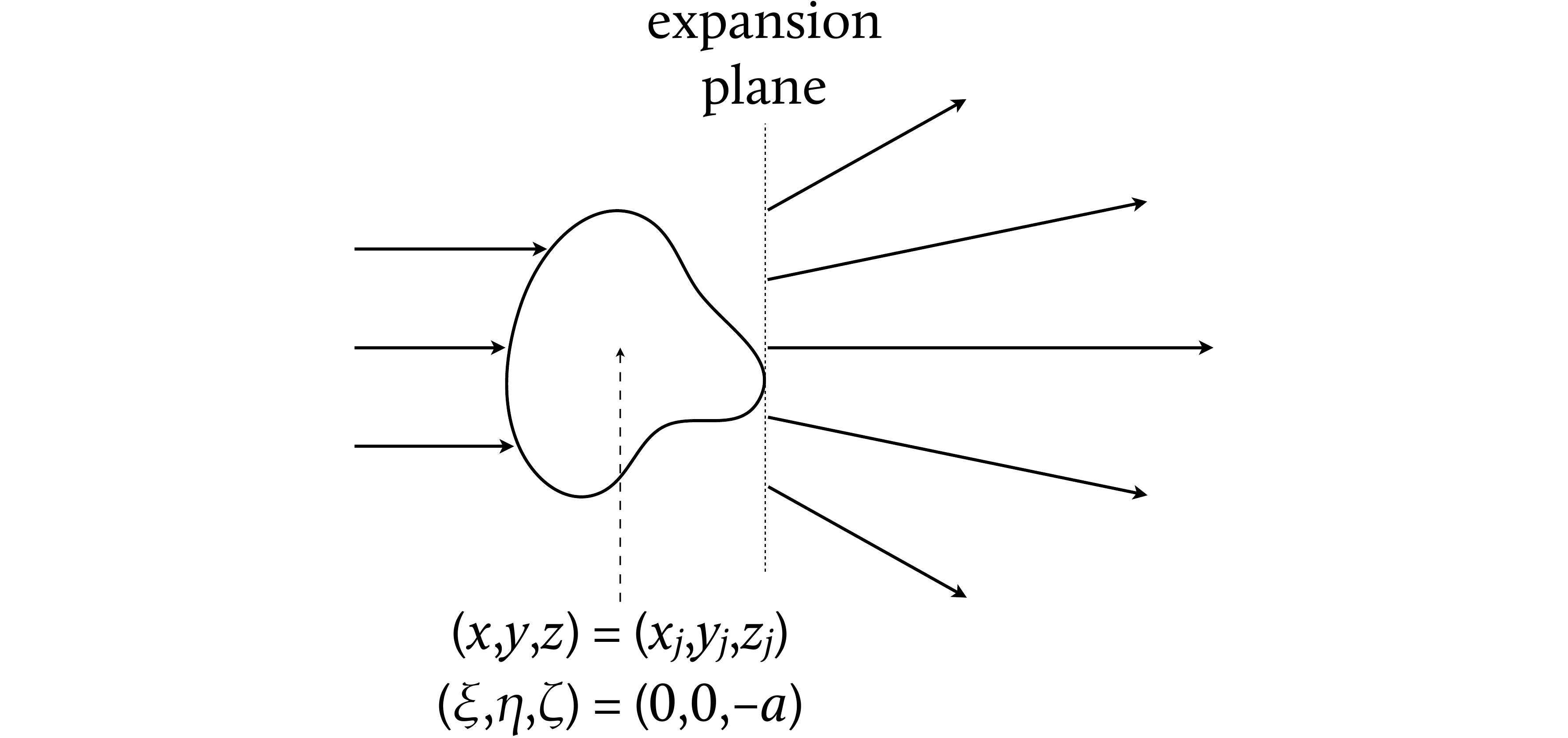}{\figwidth}{The coordinate change $\xi=x-x_j$, $\eta=y-y_j$, $\zeta=z-(z_j+a)$ allows us to calculate the scattered field as though the scattering particle were centred at $(0,0,-a)$.}{fig-FEo}

With the coordinate change $\xi=x-x_j$, $\eta=y-y_j$, $\zeta=z-(z_j+a)$, we can write
\eq{\FEs_j(\kx,\ky,\zeta=0) & = & \frac{e^{-i(\kx x_j+\ky y_j)}}{2\pi}\ \times\nonumber\\
& & \times\integral{-\infty}{\infty}{\integral{-\infty}{\infty}{\Es_j(\xi,\eta,\zeta=0)\ \times}{}}{}\nonumber\\
& & \phantom{\times\integral{-\infty}{\infty}{\integral{-\infty}{\infty}{}{}}{}}\times e^{-i(\kx\xi+\ky\eta)}\,\diff{\xi}\,\diff{\eta}.\nonumber\\}
This allows us to calculate $\Es_j$ in terms of the field that would be scattered by a particle centred at $(0,0,-a)$ with expansion plane $\zeta=0$ (figure~\ref{fig-FEo}). We write $\Es_j(x,y,z_j+a)$ in terms of the new coordinates as
\eq{\Es_j(\xi+x_j,\eta+y_j,\zeta=0) & = & \Eso(\xi,\eta,\zeta=0)\times\nonumber\\
& & \times\ e^{i\left(\keff(z_j-a)+2\km a\vphantom{b^2}\right)}\hat{e},\nonumber\\\label{eq-Es00a}}
whereby
\eq{\FEs_j(\kx,\ky,\zeta=0) & = & \FEo(\kx,\ky)\times\nonumber\\
& & \times\ e^{i\left(-(\kx x_j+\ky y_j)+\keff(z_j-a)+2\km a\vphantom{b^2}\right)}.\nonumber\\\label{eq-FEs00a}}
Note that the right side of equation~\ref{eq-Es00a} has no dependence on the transverse components of $\vec{r}_j$ (i.e.~$x_j$ and $y_j$) and that \smash{$\FEo$} in equation~\ref{eq-FEs00a} has no subindex, since it is the amplitude of the Fourier transform of the field scattered by a particle at $(0,0,-a)$, not specifically by the $j$-th particle. Note that in all of these equations the third component of the dependency of \smash{$\Es_j$} and \smash{$\FEs_j$} is the position of the expansion plane.

\fig{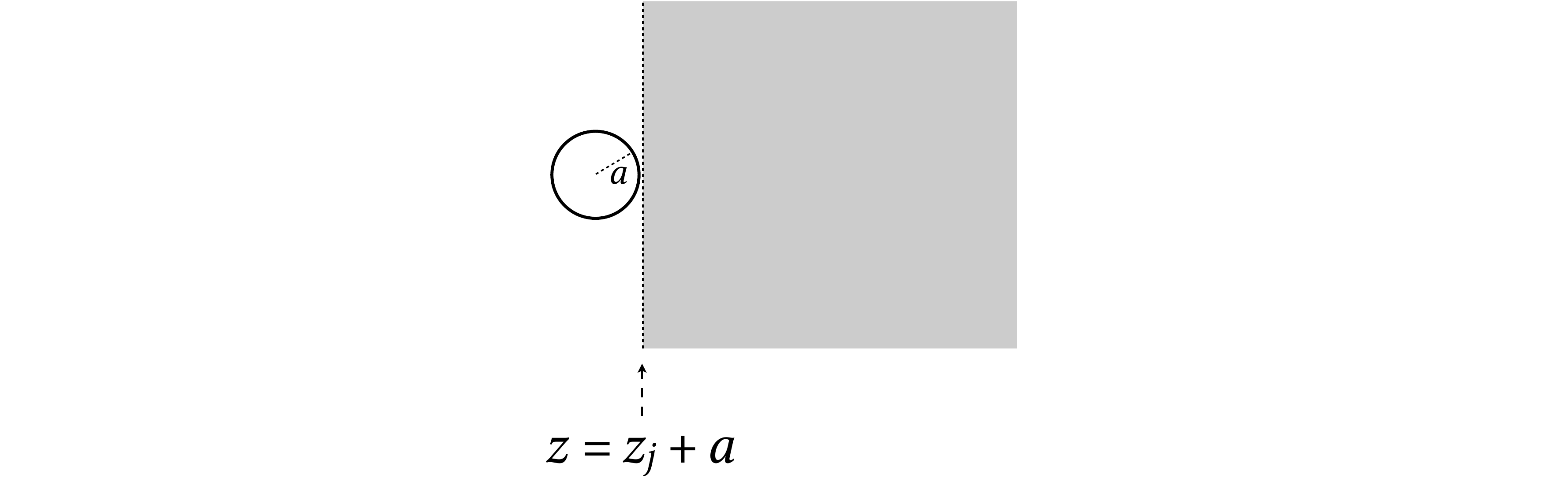}{\figwidth}{Given a position $(x,y,z)$ in the shaded region where we wish to calculate the field scattered by the $j$-th particle, equation~\ref{eq-EcAD} gives the field if the particle is located in the half-space given by $z_j\leq z-a$ (white region).}{fig-EcAD}

In the anomalous-diffraction approximation, there is no back-scattering, so a particle only affects the field inside of it and in front of it. Therefore, the effective field seen by particles located beyond the expansion plane $z=z_j+a$ (figure~\ref{fig-EcAD}) is
\eq{\Ec\evalr & = & \Ei\evalr\nonumber\\
& & +\ \frac{\rho}{2\pi}\integral{V}{}{\Theta(z-z_j-a)\ \times}{}\nonumber\\
& & \phantom{+\ \frac{\rho}{2\pi}\integral{V}{}{}{}}\times\integral{-\infty}{\infty}{\integral{-\infty}{\infty}{\FEs_j(\kx,\ky,z)\ \times}{}}{}\nonumber\\
& & \phantom{+\ \frac{\rho}{2\pi}\integral{V}{}{\times\integral{-\infty}{\infty}{\integral{-\infty}{\infty}{}{}}{}}{}}\times e^{i\left(\kx x+\ky y+\km(z-z_j-a)\vphantom{b^2}\right)}\nonumber\\
& & \hphantom{+\ \frac{\rho}{2\pi}\integral{V}{}{\times\integral{-\infty}{\infty}{\integral{-\infty}{\infty}{}{}}{}}{}}\ \diff{\kx}\,\diff{\ky}\,\diff{\vec{r}_j},\label{eq-EcAD}}
where $\Theta$ is the Heaviside step function. The phase $\km(z-z_j-a)$ takes into account the distance travelled by the field before reaching the expansion plane.

Since \smash{$\FEs_j(\kx,\ky,z)$} is the only term that depends on $x_j$ and $y_j$ (see equations~\ref{eq-Es00a} and~\ref{eq-FEs00a}) and
\eq{\integral{-\infty}{\infty}{\integral{-\infty}{\infty}{e^{-i(\kx x_j+\ky y_j)}}{x_j}}{y_j} & = & 4\pi^2\delta(\kx)\delta(\ky),\nonumber\\}
equation~\ref{eq-EcAD} simplifies to
\eq{\Ec\evalr & = & \Ei\evalr\nonumber\\
& & +\ 2\pi\rho\integral{0}{z-a}{\integral{-\infty}{\infty}{\integral{-\infty}{\infty}{\FEo(\kx,\ky)e^{i\km z}\ \times}{}}{}}{}\nonumber\\
& & \phantom{+\ 2\pi\rho\integral{0}{z-a}{\integral{-\infty}{\infty}{\integral{-\infty}{\infty}{}{}}{}}{}}\times e^{i(\keff-\km)(z_j-a)}\times\nonumber\\
& & \phantom{+\ 2\pi\rho\integral{0}{z-a}{\integral{-\infty}{\infty}{\integral{-\infty}{\infty}{}{}}{}}{}}\times\delta(\kx)\delta(\ky)\nonumber\\
& & \phantom{+\ 2\pi\rho\integral{0}{z-a}{\integral{-\infty}{\infty}{\integral{-\infty}{\infty}{}{}}{}}{}}\ \diff{\kx}\,\diff{\ky}\,\diff{z_j}\nonumber\\
& = & \Ei\evalr+2\pi\rho\,\FEo(0,0)e^{i\left(-\keff a+\km(z+a)\vphantom{b^2}\right)}\times\vphantom{\frac{b^2}{b^2}}\nonumber\\
& & \phantom{\Ei\evalr+}\ \times\frac{e^{i(\keff-\km)(z-a)}-1}{i(\keff-\km)}\,.\label{eq-Ec}}

Recall that $\Ei=\Eio e^{i\km z}\hat{e}$, $\Ec=\Eco e^{i\keff z}\hat{e}$ and, using the effective-field and anomalous-diffraction approximations,
\eq{\FEo(\kx,\ky) & = & \frac{1}{2\pi}\integral{-\infty}{\infty}{\integral{-\infty}{\infty}{\left(e^{i(\kp-\km)b}-1\right)\!\ \times}{}}{}\nonumber\\
& & \phantom{\frac{1}{2\pi}\integral{-\infty}{\infty}{\integral{-\infty}{\infty}{}{}}{}}\times\Eco e^{-i(\kx x+\ky y)}\,\diff{x}\,\diff{y}\,\hat{e}\nonumber\\}
($\Eco$ in the integral should, if the effective-field approximation is not made, be replaced by \smash{$\langle\Eeo\rangle_j$}, where \smash{$\langle\Ee\evalr\rangle_j=\langle\Eeo e^{i\keff z}\rangle_j$} and the subindex $j$ indicates that the average is taken over all particles other than the $j$-th one while keeping the $j$-th one's position fixed). Therefore, in order for equation~\ref{eq-Ec} to hold for every $z$, the equalities
\eq{\Eco e^{i\keff z} & = & \frac{\rho\,\Eco e^{i\left(\keff z-2(\keff-\km)a\vphantom{b^2}\right)}}{i(\keff-\km)}\ \times\nonumber\\
& & \times\integral{-\infty}{\infty}{\integral{-\infty}{\infty}{\left(e^{i(\kp-\km)b}-1\right)}{x}}{y}\label{eq-kez}}
and
\eq{0 & = & \Eio e^{i\km z}-\frac{\rho\,\Eco e^{i\left(\km z-(\keff-\km)a\vphantom{b^2}\right)}}{i(\keff-\km)}\ \times\nonumber\\
& & \hphantom{\Eio e^{i\km z}-}\ \times\integral{-\infty}{\infty}{\integral{-\infty}{\infty}{\left(e^{i(\kp-\km)b}-1\right)}{x}}{y}\nonumber\\}
must be satisfied; these equalities come from separating the expression for \smash{$\Eco$} into terms proportional to \smash{$e^{i\keff z}$} and terms proportional to \smash{$e^{i\km z}$}. Equation~\ref{eq-kez} simplifies to
\eq{\keff & = & \km+\frac{\rho e^{-2i(\keff-\km)a}}{i}\ \times\nonumber\\
& & \hphantom{\km+}\ \times\integral{-\infty}{\infty}{\integral{-\infty}{\infty}{\left(e^{i(\kp-\km)b}-1\right)}{x}}{y}\nonumber\\
& = & \km\!\left(1+\frac{2\pi i\rho e^{-2i(\keff-\km)a}\Sfwd}{\km\.^3}\right)\!,\label{eq-vdHpre}}
where
\eq{\Sfwd & = & \frac{\km\.^2}{2\pi}\integral{-\infty}{\infty}{\integral{-\infty}{\infty}{\left(1-e^{i(\kp-\km)b}\right)}{x}}{y}\label{eq-S}}
is the forward-scattering amplitude of a particle\cite{ref-vdH-LSSP}.

Since the particles are tenuous, we may expect $\keff$ not to be too dissimilar to $\km$: $e^{-2i(\keff-\km)a}\approx1$. We thus simplify the above expression to
\eq{\keff & \approx & \km\!\left(1+\frac{2\pi i\rho\Sfwd}{\km\.^3}\right)\!,\label{eq-vdH}}
which is van de Hulst's formula\cite{ref-vdH-LSSP}.

A first correction would be to substitute equation~\ref{eq-vdH} on the right-hand side of equation~\ref{eq-vdHpre}, yielding
\eq{\keff^{(2)} & = & \km\!\left(1+\frac{2\pi i\rho e^{4\pi\rho a\Sfwd/\km\.^2}\Sfwd}{\km\.^3}\right)\!.\label{eq-keff2}}
We therefore have a condition for van de Hulst's formula to be valid: that $|\Delta|\ll1$ with
\eq{\Delta & = & \frac{4\pi\rho a\Sfwd}{\km\.^2}\,.\label{eq-D}}

We can rewrite this quantity as
\eq{\Delta & \propto & \frac{4\pi f\Sfwd}{\pi a^2/\lambda_\text{m}\.^2}\ =\ \frac{3f\Sfwd}{\Xm\.^2}\,,\label{eq-Dalt}}
where $f=\rho v$ is the particle volume fraction (the fraction of the suspension's volume that is made up by the particles), $v\propto4\pi a^3/3$ is the volume of a particle (the proportionality here and in equation~\ref{eq-Dalt} becomes an equality for spherical particles), $\Xm=\km a$ is the size parameter (occasionally called the diffraction parameter), and $\lambda_\text{m}=\lambda/\nm$ is the wavelength in the medium surrounding the particles. This form of $\Delta$ makes it evident that it is a measure of the amount of light scattered (in the forward direction) per unit geometric cross-section of a particle (normalised to the wavelength); the larger the value of $f$ for a given particle volume, the more scatterers there are and the more the light is scattered; similarly, the larger the value of $\Sfwd$ for a given number of particles of a given size, the more strongly they scatter light in the forward direction and the more light is scattered.

Note that our derivation of van de Hulst's formula does not require far-field observation or a dilute suspension. This has been a source of confusion ever since the formula was first published. The presence of the particles' forward-scattering amplitude in equation~\ref{eq-vdH} is commonly thought to signify that the van de Hulst formula is only valid in the far field, but, in fact, $\Sfwd$ comes from taking the average of the field; nowhere in our derivation of equation~\ref{eq-vdH} do we ask that the point of observation be far from the particles or that the particles be in each other's far fields. As for the other point of contention, requiring a small value of $|\Delta|$ is not equivalent to demanding that $\rho$ itself be small, though a sufficiently small value of $\rho$ will certainly ensure this.

\subsection{Semi-periodic suspensions}\label{subsec-vdHsemiperiodic}

\fig{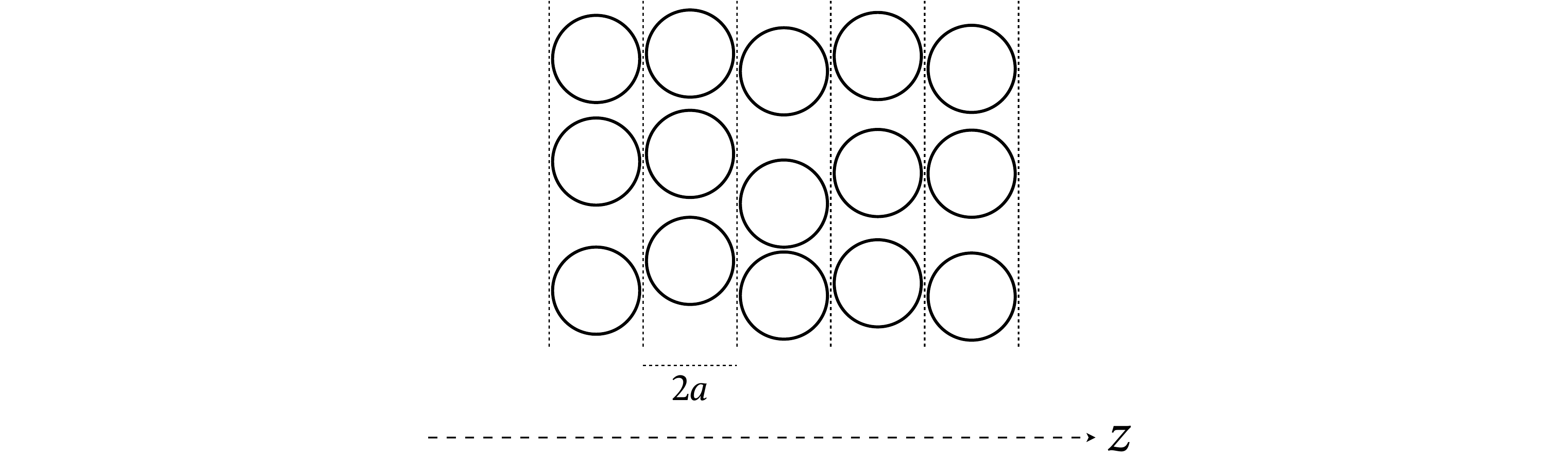}{\figwidth}{A suspension periodic in $z$ but random in $x$ and $y$ is an adequate model for epithelial tissue.}{fig-semiperiodic}

Certain kinds of tissue, such as epithelial tissue, are well approximated by stacked cellular monolayers (figure~\ref{fig-semiperiodic}). Each monolayer is one cell thick, and the cells within it can be random or ordered. The tissue is thus a semi-periodic suspension: periodic in one direction (which we will take to be the $z$ direction, in which the light propagates) but not necessarily in the perpendicular directions (those within the plane of a monolayer). van de Hulst's formula is not generally derived for this type of material, but we will do so here in order to show it is also appropriate for biological tissue. As in section~\ref{subsec-vdHrandom}, we will address the case in which the suspension is formed of only one particle type.

Consider a monolayer with front and back edges given by $z_1$ and $z_2=z_1+2a$ and containing $N$ particles. At $z_2$, the field scattered by the $j$-th particle in the monolayer is
\eq{\Es_j(x,y,z_2) & = & \left(e^{i(\kp-\km)b_j}-1\right)\!\Ei(x,y,z_2)}
(recall equation~\ref{eq-EsAD} and note that here the exciting field is $\Ei$), where $b_j(x,y)$ is the particle's thickness at $(x,y)$ (see figure~\ref{fig-Es}).

If all the particles are identical (and identically oriented), we may write $b_j(x,y)=b(x+x_j,y+y_j)$, where $b(x,y)$ is the thickness at $(x,y)$ of a particle centred at $(0,0)$, and thus
\eq{\Es(x,y,z_2) & = & \sum_{j=1}^N\left(e^{i(\kp-\km)b(x+x_j,y+y_j)}-1\right)\!\Ei(x,y,z_2).\nonumber\\}
Once again, we perform a plane-wave expansion of the field scattered by the $j$-th particle in order to calculate it beyond $z_2$:
\eq{\Es_j(x,y,z\geq z_2) & = & \frac{1}{2\pi}\integral{-\infty}{\infty}{\integral{-\infty}{\infty}{\FEs_j(\kx,\ky,z_2)\ \times}{}}{}\nonumber\\
& & \phantom{\frac{1}{2\pi}\integral{-\infty}{\infty}{\integral{-\infty}{\infty}{}{}}{}}\times e^{i(\kx x+\ky y)}\times\nonumber\\
& & \phantom{\frac{1}{2\pi}\integral{-\infty}{\infty}{\integral{-\infty}{\infty}{}{}}{}}\times e^{i\km(z-z_2)}\,\diff{\kx}\,\diff{\ky}\nonumber\\}
with
\eq{\FEs_j(\kx,\ky,z_2) & = & \frac{1}{2\pi}\integral{-\infty}{\infty}{\integral{-\infty}{\infty}{\Es_j(x,y,z_2)\ \times}{}}{}\nonumber\\
& & \phantom{\frac{1}{2\pi}\integral{-\infty}{\infty}{\integral{-\infty}{\infty}{}{}}{}}\times e^{-i(\kx x+\ky y)}\,\diff{x}\,\diff{y}.\nonumber\\}

We stress here the fact that this is not an exact representation of epithelial tissue. In real tissue, cells are not constrained to stay entirely within the bounds of a flat region of thickness $2a$ in the $z$ direction; cells are not identical to each other, and cells from one layer will partially fit themselves into the spaces between adjacent cells in the layer underneath. However, it allows us to perform simple calculations and it is not unreasonable as an approximation for tissue consisting of a stack of cellular layers; any results we obtain here are likely to be correct to within an order of magnitude for real tissue.

Using the coordinate change $\xi=x+x_j$, $\eta=y+y_j$, the Fourier transform of the total scattered field becomes
\eq{\FEs(\kx,\ky,z_2) & = & \frac{\Eio e^{i\km z_2}}{2\pi}\integral{-\infty}{\infty}{\integral{-\infty}{\infty}{\left(e^{i(\kp-\km)b(\xi,\eta)}\right.}{}}{}\nonumber\\
& & \hphantom{\frac{\Eio e^{i\km z_2}}{2\pi}\integral{-\infty}{\infty}{\integral{-\infty}{\infty}{}{}}{}}\left.\vphantom{e^{i(\kp-\km)b(\xi,\eta)}}-1\right)e^{-i(\kx\xi+\ky\eta)}\vphantom{\integral{}{}{e^{i(\kp-\km)b(\xi,\eta)}}{}}\nonumber\\
& & \hphantom{\frac{\Eio e^{i\km z_2}}{2\pi}\integral{-\infty}{\infty}{\integral{-\infty}{\infty}{}{}}{}}\ \diff{\xi}\,\diff{\eta}\times\vphantom{\frac{b}{b}}\nonumber\\
& & \times\sum_{j=1}^Ne^{i(\kx x_j+\ky y_j)}\,\hat{e}.}

Suppose now that the particles' two-dimensional positions, while random, are fixed within the monolayer. We must average over all possible translations of the monolayer within the $xy$ plane; since the particles are fixed with respect to each other, this is equivalent to integrating over all the possible two-dimensional positions $\vec{r}_j^\perp$ of the $j$-th particle with the understanding that the entire monolayer moves rigidly with this particle. The translation vector (or, equivalently, the particle's position) is governed by a uniform probability density $1/A$ within a region of area $A$, since there is nothing making one translation more likely than another. The average scattered field at $z\geq z_2$ is then
\eq{\langle\Es\evalr\rangle & = & \frac{e^{i\km(z-z_2)}}{2\pi A}\integral{A}{}{\integral{-\infty}{\infty}{\integral{-\infty}{\infty}{\FEs(\kx,\ky,z_2)\ \times}{}}{}}{}\nonumber\\
& & \hphantom{\frac{e^{i\km(z-z_2)}}{2\pi A}\integral{A}{}{\integral{-\infty}{\infty}{\integral{-\infty}{\infty}{}{}}{}}{}}\times e^{i(\kx x+\ky y)}\vphantom{\frac{b}{b}}\nonumber\\
& & \hphantom{\frac{e^{i\km(z-z_2)}}{2\pi A}\integral{A}{}{\integral{-\infty}{\infty}{\integral{-\infty}{\infty}{}{}}{}}{}}\ \diff{\kx}\,\diff{\ky}\,\diff{\vec{r}_j^\perp}\vphantom{\frac{a}{a}}\nonumber\\
& = & \frac{2\pi}{A}\integral{-\infty}{\infty}{\integral{-\infty}{\infty}{\FEs(\kx,\ky,z_2)\ \times}{}}{}\nonumber\\
& & \times\ e^{i\left(\kx x+\ky y+\km(z-z_2)\vphantom{b^2}\right)}\,\delta(\kx)\,\delta(\ky)\,\diff{\kx}\,\diff{\ky}\vphantom{\frac{b}{b}}\nonumber\\
& = & \frac{2\pi\FEs(0,0,z_2)e^{i\km(z-z_2)}}{A}\,,}
where the second equality is due to the fact that the only part that depends on $\vec{r}_j^\perp$ is the sum of exponentials.

Note that
\eq{\FEs(0,0,z_2) & = & \frac{\Eio e^{i\km z_2}}{2\pi}\integral{-\infty}{\infty}{\integral{-\infty}{\infty}{\left(e^{i(\kp-\km)b(x,y)}\right.}{}}{}\nonumber\\
& & \hphantom{\frac{\Eio e^{i\km z_2}}{2\pi}\integral{-\infty}{\infty}{\integral{-\infty}{\infty}{}{}}{}}\left.\vphantom{e^{i(\kp-\km)b(x,y)}}-1\right)\diff{x}\,\diff{y}\times\vphantom{\integral{}{}{e^{i(\kp-\km)b(x,y)}}{}}\nonumber\\
& & \times\sum_{j=1}^N1\,\hat{e}\nonumber\\
& = & -\frac{N\Eio e^{i\km z_2}\Sfwd}{\km\.^2}\,\hat{e},}
where $\Sfwd$ is given by equation~\ref{eq-S}. Thus,
\eq{\langle\Es\evalr\rangle & = & -\frac{4\pi\rho a\Sfwd}{\km\.^2}\,\Ei\evalr,}
where $\rho=N/2aA$ is the density of particles in the monolayer.

The average field at $z_2$ is
\eq{\Ec(x,y,z_2) & = & \Ei(x,y,z_2)+\langle\Es(x,y,z_2)\rangle\nonumber\\
& = & \Ei(x,y,z_2)\left(1-\frac{4\pi\rho a\Sfwd}{\km\.^2}\right)\!.\label{eq-Ecmono}}
If there is now a stack of $M$ identical monolayers (identical in the type and density of particles, though not necessarily in the exact positions of the particles), then the field of equation~\ref{eq-Ecmono} is incident on the second monolayer, after which the field is given by
\eq{\Ec\evalr & = & \Ei\evalr\left(1-\frac{4\pi\rho a\Sfwd}{\km\.^2}\right)^2,}
and so on; at the end of the stack, the field is
\eq{\Ec\evalr & = & \Ei\evalr\left(1-\frac{4\pi\rho a\Sfwd}{\km\.^2}\right)^M\nonumber\\
& = & \Ei\evalr\left(1-\frac{2\pi\rho D\Sfwd}{M\km\.^2}\right)^M,}
where $D=2aM$ is the thickness of the stack. We rewrite this as
\eq{\Ec\evalr & = & \Ei\evalr\left(\left(1-\frac{2\pi\rho D\Sfwd}{M\km\.^2}\right)^{-\frac{M\km\.^2}{2\pi\rho D\Sfwd}}\right)^{-\frac{2\pi\rho D\Sfwd}{\km\.^2}}.\nonumber\\\label{eq-Ecpowers}}
If the quantity $2\pi\rho D|\Sfwd|/M\km\.^2$ is very small, then
\eq{\Ec\evalr & \approx & \Ei\evalr e^{-2\pi\rho D\Sfwd/\km\.^2}\nonumber\\
& = & \Eio e^{i\km(z-D)+i\km\left(1+2\pi i\rho\Sfwd/\km\.^3\vphantom{b^2}\right)D}\hat{e},\label{eq-Ecstack}}
where we have used the definition
\eq{e & \equiv & \limit{q\to0}(1+q)^{1/q}.}

Comparing equation~\ref{eq-Ecstack} to the expression for a field that has gone through a medium of thickness $D$ with propagation constant $\keff$ immersed in a medium with propagation constant $\km$,
\eq{\vec{E}\evalr & = & \Eio e^{i\km(z-D)+i\keff D}\hat{e},}
we finally obtain an expression for the effective propagation constant of the stack of monolayers:
\eq{\keff & \approx & \km\!\left(1+\frac{2\pi i\rho\Sfwd}{\km\.^3}\right)\!,}
which is again van de Hulst's formula.

It is perhaps surprising that a partially ordered, dense (especially in the $z$ direction) medium is accurately described by van de Hulst's formula, which was conceived for completely random media. Once again, however, we do not require a dilute suspension or that the particles be in each other's far fields (here, in fact, they are very much \emph{not} in each other's far fields); we merely require that $2\pi\rho D|\Sfwd|/M\km\.^2\equiv4\pi\rho a|\Sfwd|/\km\.^2\equiv|\Delta|$ be small, as in the case of the random medium.

\section{Validity of the van de Hulst formula for biological suspensions}\label{sec-vdHvalidity}

We have derived van de Hulst's formula for a random medium and a semi-periodic medium and found the limiting factor for the formula's validity to be the dimensionless quantity $\Delta$ (or, more accurately, its magnitude; see equation~\ref{eq-D}) in both cases. The formula is
\eq{\keff & \approx & \km\!\left(1+\frac{2\pi i\rho\Sfwd}{\km\.^3}\right)}
for the effective propagation constant or, equivalently,
\eq{\neff & \approx & \nm\!\left(1+\frac{2\pi i\rho\Sfwd}{\km\.^3}\right)\ \approx\ \nm\!\left(1+\frac{3if\Sfwd}{2\Xm\.^3}\right)\nonumber\\\label{eq-vdHn}}
for the effective refractive index, where the approximation symbol becomes an equals sign for spheres. For the formula to hold, we require $|\Delta|\ll1$, which physically means that the amount of forward-scattered light per (normalised) particle area must be low (see discussion at the end of section~\ref{subsec-vdHrandom}). Note that this is a stronger constraint than requiring that $3f|\Sfwd|/2\Xm\.^3\ll1$ (i.e.~that the magnitude of the second term in brackets in equation~\ref{eq-vdHn} be small), as the latter term contains an additional $\Xm$ in the denominator and for the systems under consideration here $\Xm\gg1$.

We will now look at whether $|\Delta|$ can indeed be very small for realistic biological suspensions.

\begin{figure}[b!]
\begin{center}
\begin{tikzpicture}
\pgfplotsset{every axis legend/.append style={at={(0.8,0.11)},anchor=west}}
\begin{axis}[width=\graphwidth,height=\graphwidth,legend cell align=left,legend style={draw=none},xlabel=$\lambda$ (nm),xmin=400,xmax=1025,ymin=-0.17,ymax=0.28,scaled ticks=false,yticklabel style={/pgf/number format/.cd,fixed,fixed zerofill,precision=2},axis x line=middle,x label style={at={(ticklabel* cs:1)},anchor=west},axis y line=left,y label style={at={(current axis.above origin)},anchor=south}]
\addplot[smooth,thick,black,dashed] file[x index=0,y index=1] {fig-ReD.dat};
\addplot[smooth,thick,black,dotted] file[x index=0,y index=1] {fig-ImD.dat};
\addplot[smooth,thick,black] file[x index=0,y index=1] {fig-AbsD.dat};
\legend{$\Real{\Delta}$,$\Imag{\Delta}$,$|\Delta|$}
\end{axis}
\end{tikzpicture}
\captionsetup{singlelinecheck=off}
\end{center}
\caption[.]{$\Delta$ for a suspension of spheres with a radius of 2.65~\um\ and the refractive index of an erythrocyte submerged in a medium with the refractive index of blood plasma.}
\label{fig-D}
\end{figure}
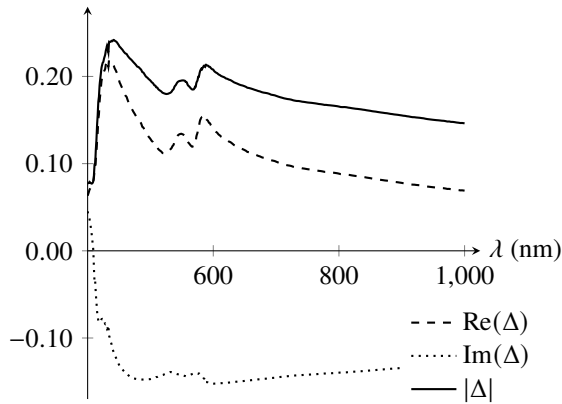

An average erythrocyte has a volume close to 78~\ummm\cite{ref-NahmadRohen-PS91} and a refractive index close to $1.42+(10^{-4}\text{--}10^{-2})i$ in the visible range of the spectrum\cite{ref-Borovoi-PSPIE3194}. Blood plasma has a refractive index of about $1.35$ across this wavelength range\cite{ref-Reddy-IJITCE2}. Though an erythrocyte is a biconcave disc whose shape is accurately described by a quadratic equation with cylindrical symmetry and no odd-powered terms\cite{ref-NahmadRohen-PS91,ref-Kuchel-BMB61}, for the sake of order-of-magnitude calculations we may use spheres; the radius of a sphere with a volume of 78~\ummm\ is about 2.65~\um. Figure~\ref{fig-D} shows $\Delta$ for a suspension of spheres with the refractive index of an erythrocyte submerged in a medium with the refractive index of blood plasma (using a volume fraction of 0.45) for wavelengths from 400~nm to 1,000~nm; results should be similar for other cell types, as the refractive-index contrast is very similar\cite{ref-Xu-OL30}. $|\Delta|/\rho$ varies between 125~\ummm\ and 175~\ummm. Dense packing of spheres this size gives $\rho\approx9.5\times10^{-3}$~\mmmu, which yields $|\Delta|\sim0.5$. Now, this is not much less than 1, but in blood, where the haematocrit (the fraction of the blood's volume that is made up of the erythrocytes) is 0.40--0.45\cite{ref-Guyton-TMP}, $\rho$ becomes about $5\times10^{-3}$~\mmmu~$=5\times10^6$~$\upmu$l$^{-1}$, which gives $|\Delta|\sim0.2$. The relative error incurred by making the approximation $e^{|\Delta|}\approx1$ (compare equations~\ref{eq-vdH} and~\ref{eq-keff2}) is $1-1/e^{|\Delta|}\approx20\%$, meaning the van de Hulst formula underestimates the particles' contribution (i.e.~the quantity $\keff-\km$) by one fifth. Given the magnitude of this contribution in the case of blood\cite{ref-NahmadRohen-PS91}, the error in the effective refractive index is only about $(1\text{--}3)\times10^{-3}+(2\text{--}5)\times10^{-3}\,i$ in this case.

\begin{figure}[t!]
\begin{center}
\begin{tikzpicture}
\pgfplotsset{every axis legend/.append style={at={(0.8,0.48)},anchor=west}}
\begin{axis}[width=\graphwidth,height=\graphwidth,legend cell align=left,legend style={draw=none},xlabel=$\lambda$ (nm),xmin=400,xmax=1025,ymin=-0.13,ymax=1.25,scaled ticks=false,yticklabel style={/pgf/number format/.cd,fixed,fixed zerofill,precision=1},axis x line=middle,x label style={at={(ticklabel* cs:1)},anchor=west},axis y line=left,y label style={at={(current axis.above origin)},anchor=south}]
\addplot[smooth,thick,black,dashed] file[x index=0,y index=1] {fig-Reg.dat};
\addplot[smooth,thick,black,dotted] file[x index=0,y index=1] {fig-Img.dat};
\addplot[smooth,thick,black] file[x index=0,y index=1] {fig-Absg.dat};
\legend{$\Real{g}$,$\Imag{g}$,$|g|$}
\end{axis}
\end{tikzpicture}
\captionsetup{singlelinecheck=off}
\end{center}
\caption[.]{The quantity $g$, which is a measure of the error of van de Hulst's formula, for a suspension of $5\times10^{6}$~$\upmu$l$^{-1}$ spheres; the spheres have a radius of 2.65~\um\ and the refractive index of an erythrocyte, while the surrounding medium has the refractive index of blood plasma.}
\label{fig-g}
\end{figure}
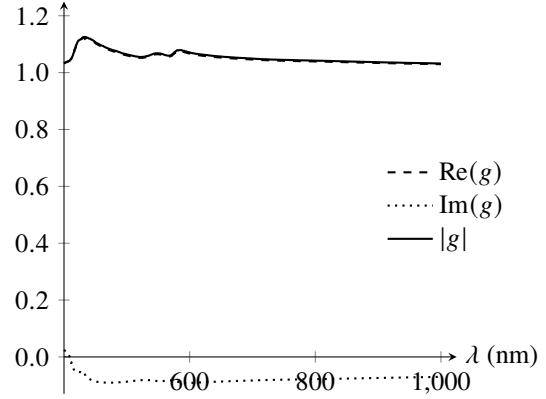

The derivation of the formula for a semi-periodic suspension gives us an alternative way to calculate this error. To obtain equation~\ref{eq-Ecstack} from equation~\ref{eq-Ecpowers}, we have assumed $|\Delta|$ is very small. If we write
\eq{G & = & (1-\Delta)^{-1/\Delta}}
and
\eq{g & = & \log{G}}
(figure~\ref{fig-g}), the average field after a stack of $M$ monolayers becomes
\eq{\Ec\evalr & = & \Eio e^{i\km(z-D)+i\km\left(1+2\pi i\rho g\Sfwd/\km\.^3\vphantom{b^2}\right)D}\hat{e},}
which reduces to equation~\ref{eq-Ecstack} when $g=1$. Here, $1-1/g$ is the relative error. For $\rho=5\times10^{-3}$~\mmmu\ and the same refractive indices and sphere radius as before, we have $g=(1.00\text{--}1.15)-(0.00\text{--}0.10)i$, which gives an error of 13\% or less throughout the (400--1,000)-nm range. Furthermore, the simple modification
\eq{\keff & \approx & \km\!\left(1+\frac{2\pi i\rho g\Sfwd}{\km\.^3}\right)}
yields the correct effective propagation constant.

Anomalous diffraction itself is an approximation, and therefore its use might raise concerns about the error incurred by it (compared to Mie theory, for example, which is exact). In a previous study\cite{ref-NahmadRohen-PS91}, we found that the extinction coefficient of a dilute suspension of erythrocyte-like spheres (meaning they have the same volume and refractive index as real erythrocytes) calculated with anomalous diffraction has an error of less than 8.25\% throughout the visible spectrum and less than 5\% for most of it ($\lambda\geq550$~nm) compared to the calculation with Mie theory. For all wavelengths, the contribution of the particles is \emph{over}estimated. Therefore, this error not only is small, but partially compensates the underestimation incurred by van de Hulst's formula itself.

\section{Discussion \& conclusions}

The van de Hulst formula gives a simple expression (though not necessarily one easy to compute for complicated particle shapes\cite{ref-NahmadRohen-PS91}) for the effective propagation constant $\keff$ or the effective refractive index $\neff$ of a suspension of particles. It has been widely used\cite{ref-Mohammadi-ACIS62,ref-Alexander-JCSFT277,ref-Meeten-MST8,ref-Borovoi-PSPIE3194,ref-Goulden-TFS54,ref-Champion-FT2,ref-Diehl-OA3,ref-Stramski-AO27,ref-Granovskii-ZETF105,ref-Meeten-MST6,ref-Konevskikh-FD187,ref-MoralesLuna-MST28}, yet there are some misconceptions about its requirements. The original derivation of the formula\cite{ref-vdH-LSSP} and some modern derivations\cite{ref-Bohren-ASLSP} commonly look at the phase shift introduced in the coherent field by a thin slab filled with the suspension in question and employ anomalous diffraction to calculate the forward-scattering amplitude $\Sfwd$, but these derivations, as well as more-classical ones\cite{ref-Foldy-PR67,ref-Lax-RMP23,ref-Twersky-JOSA52,ref-Twersky-JMP3,ref-Twersky-JOSA60a}, invariably require that the particles be in each other's far fields, that the point of observation be in the far field of the slab and/or that the sample be dilute. In particular, Twersky's derivation of a van-de-Hulst-like formula further requires that the suspended particles be either smaller than the wavelength or so tenuous that the refractive index contrast is less than 0.01. Other derivations\cite{ref-Barrera-JOSAA20,ref-Barrera-PRB75,ref-GutierrezReyes-PSSb249} are free of far-field restrictions but do require extremely dilute suspensions.

We have derived van de Hulst's formula for both a suspension of randomly positioned particles (a random medium) and a stack of monolayers of laterally randomly positioned particles (a semi-periodic medium). In both cases, we took the effective-field and anomalous-diffraction approximations as a starting point. Our derivations show that, contrary to popular belief, the formula does not necessarily become invalid when the density of particles $\rho$ is high or the point of observation is not in the far field; nowhere do we impose these restrictions.

Conversely, we have shown that the only true restriction for biological suspensions is that the quantity $\Delta=4\pi\rho a\Sfwd/\km\.^2$ be small. We have also shown two ways of computing and correcting the error incurred by van de Hulst's formula in the case in which $|\Delta|$ is not much smaller than 1; each of these ways is suggested by the very derivation of the formula for one of the suspension types studied. The requirement $f\ll1$ is not satisfied for biological suspensions such as blood and epithelial tissue, yet we have shown that van de Hulst's formula gives surprisingly accurate estimations of $\keff-\km$ or $\neff-\nm$, the particles' contribution to the effective propagation constant or the effective refractive index, respectively, for these suspensions.

While on the subject of cells' contribution to the effective properties of a biological suspension they are part of, we note that it is important for the suspension to be modelled taking into account the effects of scattering by cells, which are much larger than the wavelength of visible light, whether the model uses the van de Hulst formula or some other expression for the calculation. This was recognised by Twersky\cite{ref-Twersky-JOSA60b} five decades ago, yet many recent studies\cite{ref-Li-PSPIE3914,ref-Liu-JBO24} have still employed haemolysed blood, which is essentially a haemoglobin solution without any scattering cells, as an approximation for whole blood. Since a haemoglobin solution contains no scatterers, it cannot be an accurate representation of whole blood as an effective medium; if it were, measurements of the refractive index throughout a cytolysis process would reveal no variation during the process, which is not the case\cite{ref-MarquezIslas-MST31}. We believe the present derivation of an appropriate scattering model and our previous work on the subject\cite{ref-NahmadRohen-PS91} clarify why calculating the effective propagation constant of biological suspensions using haemolysed blood yields results which are not representative of blood in its biological state.

Recently Rowe et al\cite{ref-Rowe-SR7} measured the refractive index of whole blood by reflection for wavelengths up to 15,000~nm. It would be interesting to calculate the effective refractive index of whole blood up to such wavelengths using van de Hulst's formula and compare the results to Rowe's; this would be a good indicator of how far the formula can be pushed, since in the infrared region of the electromagnetic spectrum the particle size is no longer much larger than the wavelength and thus anomalous diffraction might cease to be an adequate approximation.

Our derivation of van de Hulst's formula is, of course, for suspensions which are characterised by a large particle size and a low optical contrast and are thus amenable to the anomalous-diffraction approximation. Other types of suspension might be accurately described by the van de Hulst formula or similar expressions only on the condition that $\rho$ be small and/or only for far-field observation. Likewise, there may well be suspensions for which anomalous diffraction is not applicable and yet van de Hulst's formula is appropriate (meaning $|\Delta|\ll1$ or $g\approx1$) without invoking either small $\rho$ or far-field observation; this will be addressed in future publications.

In the future, it would be interesting to discuss the relationship between van de Hulst's $\keff$ for the coherent component of light and the radiative-transfer equation. We also leave this for another publication.

\section{Disclosures}

The authors declare no conflicts of interest.

\section{References}

\begingroup
\renewcommand{\section}[2]{}

\endgroup

\vfill

\end{document}